\documentclass[12pt]{article} 
\pdfoutput=1
\usepackage{psfig}
\usepackage{pstricks}
\usepackage{graphicx}
\usepackage{amsmath}

\begin{document}

\begin{center}
{\Large \bf{
Determination of the damping coefficient of electrons in optically transparent glasses at the true resonance frequency in the ultraviolet from an analysis of the Lorentz-Maxwell model of dispersion}}
\end{center}

\begin{center}
{Surajit Chakrabarti\\(Ramakrishna Mission Vidyamandira)\\  Howrah, India
}
\end{center}


\noindent
The Lorentz-Maxwell  model of dispersion of light  has been analyzed  in this paper  to determine the true  resonance frequency in the ultraviolet for the electrons in  optically transparent  glasses and the damping coefficient at this frequency.  For this we needed the refractive indices of glass in the optical frequency range. We argue that  the true resonance condition in the absorption region prevails when the  frequency at which the absorption coefficient is maximum is the same as the frequency at which the average energy per cycle of the electrons is also a maximum.  We have simultaneously  solved  the two equations obtained from the two maxima conditions numerically to arrive at a unique solution for the true resonance frequency and the damping coefficient at this frequency. Assuming  the damping coefficient to be constant over a small frequency range in the absorption region, we have determined the frequencies at which the extinction coefficient and the reflectance are maxima. These frequencies  match very well with the published data for silica glasses available from the literature.



\newpage
\section{Introduction}
\noindent
The Lorentz-Maxwell model of dispersion of electromagnetic waves in matter  is very successful  in describing the properties of matter under the action of electromagnetic waves over its  whole spectrum where the wavelength is large compared to the interatomic distances. The  model is generally studied in the optical frequency range where only the oscillation of electrons bound to atoms and molecules is relevant for the study of dispersion. Two important parameters of the model namely the natural oscillation frequency and the plasma frequency of the electrons  in a dielectric medium like glass can be easily determined from the refractive indices of a glass prism measured in the optical band [1] where glass is transparent. In a condensed system like glass one has to include the effect of the  local field on the electrons apart from the field of the incident wave. This leads to another frequency which is conventionally known as the resonance frequency and is related to the plasma and the natural oscillation frequencies of the electron [2]. Though it is called the resonance frequency, there is no proof  that the absorption coefficient  is maximum at this frequency.

   In order to study the absorption of EM waves in matter, a phenomenological variable called the damping coefficient  is introduced in the Lorentz-Maxwell model. Glass is  opaque in the ultraviolet indicating that it has a strong absorption there.  In scientific literature, there are  innumerable experimental works which have studied  the  interaction of silica glasses with electromagnetic waves  over its whole  spectrum. A summary of these works can be found in Kitamura et al. [3]. From the experimental data on the extinction coefficient for silica glass in the ultraviolet, we can find  the frequency at which this coefficient is maximum. However, as far as we are aware, there has  been no  theoretical study so far  which has determined this frequency  by an analysis of the Lorentz-Maxwell model of dispersion. The main problem with the theoretical analysis  is the fact that  it has not been possible so far to determine the value of the damping coefficient theoretically. 

In this work we have  determined  the damping coefficient at the true resonance frequency which we define to be the frequency at which the absorption coefficient for the energy of the electromagnetic field in the medium is maximum. We have done this theoretically by taking the natural  oscillation frequency and the plasma frequency determined from the refractive indices of glass in the optical region as two known  parameters of the Lorentz-Maxwell model. We have formed two algebraic equations containing the true resonance frequency and the damping coefficient as two unknown variables. We have solved these two equations simultaneously by numerical method to find a unique solution for the two variables. With the value of the damping coefficient known, we have explored the anomalous dispersion region in the ultraviolet for glass.
 
It is well known that the Kramers-Kronig relations [3] allow us to determine the imaginary part of the dielectric constant from an integral of the real part over the whole range of frequencies and vice versa. The theory is based on a very general causality argument  and  a linear response of the medium to an external perturbation.  We have, on the other hand, determined the damping coefficient of the Lorentz-Maxwell model of dispersion starting from the refractive indices in the optical region corresponding to the real value of the dielectric constant. From this we have extracted the information about the absorptive region in the ultraviolet corresponding to the imaginary part of the dielectric constant.

  In  section 2 we give the outline of the Lorentz-Maxwell model. In  section 3 we offer our physical argument for the method adopted to determine the damping coefficient and the true resonance frequency. The next four sections are just an execution of these ideas. We conclude with a summary of the work.

\section{Lorentz-Maxwell model of dispersion}
\noindent
In the Lorentz model [4] of dispersion of light in a dense medium like a solid or liquid, electrons execute  forced simple harmonic oscillations
with damping in the combined  field of the incident electromagnetic 
wave of frequency
$\omega$ and the local field. The local field arises as a result of the interaction of the electron with the fields of other\ atoms close by. Without any loss of 
generality we can assume that the direction
in which the electron is oscillating is the $y$ direction. We can write the 
equation of motion  as 
\begin{equation} \ddot y+\gamma \dot y +\omega_0^2 y = \frac
{qE_0^{\prime}}{m}e^{-i\omega t}.\end{equation} where $E_0^{\prime}$ is the amplitude of the  effective field acting on the electrons.
Here $\omega_0$ is the natural oscillation frequency and $\gamma$ is the damping 
coefficient of the electron. In the steady state the 
electron will oscillate at a frequency $\omega$ of the incident wave though shifted in phase.  
$E_0^{\prime}$ is related to the amplitude of the  field ($E_{i0}$)  outside from where it is incident on the medium 
as \begin{equation} E_0^{\prime}=\frac{1+\frac{\chi}{3}}{1+D \chi}E_{i0}.\end{equation}
The $\frac{\chi}{3}$ term in equation (2) arises as a result of the effect of local field in the Lorentz-Lorenz theory of dielectric polarizability valid for an isotropic medium [5] where $\chi$ is the electric susceptibility.  $D$ is the depolarisation factor,
a dimensionless number of the order of unity [6].
The dielectric function of the medium is given by \begin{equation} \epsilon=1+\chi.\end{equation}
Using Maxwell's phenomenological relation  $\epsilon=n_c^2$ where $n_c$ is the complex 
refractive index and the Lorentz-Lorenz equation [5], we arrive at the following equation for a number of resonance regions [7,8].
\begin{equation} \frac {n_c^2-1}{n_c^2+2} = \frac {Nq^2}{3\epsilon_0 m}\sum_j
\frac {f_j}{\omega_{0j}^2-\omega^2-i\gamma_j\omega}.\end{equation}Here
$f_j$ is the fraction of electrons  that have a natural 
oscillation frequency $\omega_{0j}$ and damping constant $\gamma_j$ with $\Sigma f_j=1$. $N$ is the density of electrons taking part in dispersion.
It is a common practice   to  assume a single dominant absorption frequency which is true in many practical
cases and which makes the analysis simpler [9]. With this assumption $f_j=1$ and  
 equation (4) can be written as \begin{equation} n_c^2 = 1 +\frac{\omega_p^2}{\omega_n^2
-\omega^2 - i\gamma\omega}\end{equation}  where
the plasma frequency  $\omega_p$ is given by \begin{equation} \omega_p^2 =
\frac{Nq^2}{\epsilon_0m}\end{equation}  and  we define \begin{equation} \omega_n^2=\omega_0^2-\frac{\omega_p^2}{3}.
\end{equation}
 In  scientific literature [2], $\omega_0$ is known as the natural oscillation frequency of the electrons and $\omega_n$ is known conventionally as the resonance frequency.  
So far, authors have used some chosen values of the damping coefficient $\gamma$ and the  plasma frequency which mimic the absorptive properties of dielectric materials,  in order to carry out  model analysis [9]. We have actually determined the damping coefficient  from a prior knowledge of the natural oscillation frequency and the plasma frequency of a glass medium.

In the optical limit where the absorption in glass is negligible we take $\gamma=0$. In this limit the refractive index is real and equation (5) reduces to \begin{equation} n^2 = 1 +\frac{\omega_p^2}{\omega_n^2
-\omega^2}.\end{equation} which is essentially the Sellmeier's formula [7] for  dispersion in the frequency domain  with one absorption band. If we have a set of measurements of refractive indices of a glass prism for several optical wavelengths, we can determine $\omega_n$ and $\omega_p$ using  equation (8) [1].The resonance wavelength which falls in the ultraviolet region, has been determined in a similar work [7]. Once $\omega_n$ and $\omega_p$ are known, $\omega_0$ can be determined using equation (7).

In the absorptive region the dielectric function picks up an imaginary part given by 
\begin{equation}n_c^2=\epsilon=\epsilon_1+
i\epsilon_2.\end{equation} 
The  refractive index ($n$) and  the extinction coefficient ($\kappa$) 
known as optical constants are written as
 \begin{equation}
n_c = n + i\kappa\end{equation} 
where $\kappa$ represents the attenuation factor of the amplitude of the electromagnetic wave in an absorptive medium.
Using the last two equations and equation (5) we obtain for the real and imaginary parts of the complex dielectric function [10],
\begin{equation} \epsilon_1=n^2 - \kappa^2 =
1+\frac{\omega_p^2(\omega_n^2-\omega^2)}{(\omega_n^2-\omega^2)^2
 + \gamma^2\omega^2}\end{equation} and  
\begin{equation}\epsilon_2=2n\kappa = \frac{\omega_p^2\gamma\omega}{
(\omega_n^2-\omega^2)^2 + \gamma^2\omega^2}.\end{equation} 
We can express $n^2$ and $\kappa^2$ as functions of frequency $\omega$ using 
equations (11) and (12). The details and the final expressions have been shown in Appendix A.

The absorption coefficient of the incident EM wave
$\alpha$ is given by [7,10],
\begin{equation}\alpha = \frac{2\omega}{c}\kappa\end{equation} where $c$ is the
speed of light in vacuum.  $\alpha$ gives the attenuation coefficient of the
intensity of the incident wave. Intensity is the rate of flow of energy  per unit area normal to  a surface. $\alpha$ will be a maximum at the frequency at which the  absorption of energy by the electrons from the EM field is maximum. This gives the condition of resonance. It is a general practice to  consider $\omega_n$ defined in  equation (7) as the resonance frequency though   there is no proof that the energy 
absorption is maximum at this frequency. So we do not assume a priori $\omega_n$ to be the resonance frequency.
In the next section we will describe our strategy to
find the true resonance frequency and in the results section we will see that the true resonance frequency is different from
both $\omega_0$ and $\omega_n$ and 
 lies between them. There is no real reason to call $\omega_n$ the resonance frequency. We treat the true resonance 
frequency as an unknown variable to be found  from our analysis. 

The damping coefficient $\gamma$ is introduced in the Lorentz model to explain absorption. We model $\gamma$ such that it is zero in the optical band and upto the frequency $\omega_n$. In the absorption band we assume that $\gamma$ is constant from frequency $\omega_n$ to $\omega_0$. Above $\omega_0$, $\gamma$ falls down and rises again to another constant value of $\gamma$ in the next resonance region if the material under study has one. With this model for $\gamma$ in mind we can extrapolate equation (8) to find $\omega_n$ and $\omega_p$. In the next section we will explain how to get the constant value of $\gamma$ in the absorption region and the true resonance frequency.

Even if the system under study may have  several absorption bands, we can study it with the assumption of a single resonance region.  The optical waves oscillate the outermost electrons of atoms and molecules having the lowest natural frequencies and as a result we get  the phenomenon of refraction. With an analysis of the refractive indices in the optical region under this assumption of single resonance,  we are most likely to find information about the absorption band with the lowest natural oscillation frequency in the ultraviolet closest to the optical band. This will of course depend on the strength of the resonance. The justification of the single resonance calculations with the chosen model for $\gamma$ can be found from the results of our theoretical calculations which will be found to match the experimental results very well.

\section {Physical argument for the method adopted to determine the damping 
coefficient at the true resonance frequency}
\noindent
From various experiments on the absorption of EM waves in matter, we  know that the absorption
coefficient ($\alpha$) attains  a maximum value  at a characteristic frequency. We try to find this frequency where $\alpha$ is maximum.
 We differentiate $\alpha$ with respect to frequency and equating the derivative to zero get one equation. However, we have two unknown variables in the theory - the  damping coefficient and the true resonance frequency. We look for a second equation.

 The incident EM wave interacts with the electrons bound
to the atoms and molecules.The electrons  execute a forced simple harmonic oscillation 
with damping. The total energy of the electron is time dependent, as the 
electron is being perturbed by a time dependent harmonic force.The average energy of the electron per cycle
 can be worked out easily [11]. We find the frequency at which this average energy per cycle
is maximum. This leads  to another equation involving the two unknown variables. When the frequency at which $\alpha$ is maximum is the same as the frequency at which the average energy per cycle of the electron is also a maximum, the electromagnetic wave will share its energy most with the electrons and will be attenuated most. This 
will  constitute the true condition of resonance. By solving the two equations simultaneously using  numerical method, we find both the variables. We call the characteristic frequency, the true resonance frequency $\omega_t$
and the damping coefficient at the true resonance frequency $\gamma_t$. 

 Heitler [12] has proposed a quantum theory of the phenomenon of damping. According to this theory the damping coefficient  is dependent on frequency though of a very slowly varying nature near resonances. This gives support to our earlier assumption that  the damping coefficient is  a constant within a small frequency range about the resonance frequency. However,  it can be taken as zero in the optical band where glass is transparent and absorption is negligible.

\section {Condition for the maximum of  the absorption coefficient as a function of frequency}
\noindent
 Our aim in this section is to find the frequency at which $\alpha$ is maximum. 
We first differentiate $\alpha$ with respect to $\omega$ assuming $\gamma$ constant.  In order to find the derivative of $\alpha$   we first
differentiate  equations  (11) and (12) with respect to $\omega$. We find two algebraic
equations involving $\frac{dn}{d\omega}$ and $\frac{d\kappa}{d\omega}$. By 
eliminating $\frac{dn}{d\omega}$ from the two equations, we get the expression
for $\frac{d\kappa}{d\omega}$ and hence $\frac {d\alpha}{d\omega}$ using equation (13). We have shown the differentiations in Appendix B. Eliminating
$\frac{dn}{d\omega}$ between equations (B.2) and (B.3) we get \begin{equation}2\frac
{d\kappa}{d\omega}(n+\frac{\kappa^2}{n})=\frac{A-B}{C}\end{equation} where 
\begin{equation}A=\omega_p^2
(\omega_n^2-\omega^2)^2[\gamma-2\frac{\kappa}{n}\omega]\end{equation} and 
\begin{equation}B=\omega_p^2\omega
\gamma[\omega\gamma^2-4(\omega_n^2-\omega^2)\omega-2\frac{\kappa}{n}\omega_n^2
\gamma]\end{equation} and \begin{equation} C=[(\omega_n^2-\omega^2)^2
+\gamma^2 \omega^2]^2.\end{equation}
From this we get \begin{equation}\frac{d\alpha}{d\omega}
=\frac{\kappa}{c} [2+\frac{n}{\kappa}\frac{\omega}{n^2+\kappa^2}\frac{A-B}{C}].
\end{equation}If $\alpha$ is maximum  then  $\frac{d\alpha}{d\omega}$ should 
be zero. So we write at the maximum \begin{equation}
\frac{\omega(A-B)}{C}=-2\frac{\kappa}{n}(n^2+\kappa^2).\end{equation} It is to 
be noted that two sides of equation (19) are dimensionless and they will be compared later numerically to find the 
solution for the true resonance frequency and the damping coefficient.

\section{Condition for the maximum of the average energy per cycle of the electron as a function of frequency }
\noindent
 In the steady state the electron will oscillate
at a frequency $\omega$ as given by the steady state solution of equation (1) and the total energy of the system 
averaged over a 
period is given by [11],\begin{equation} \overline {E(\omega)}=\frac{1}{4}
\frac
{(qE_0^{\prime})^2}{m}\frac{(\omega^2+\omega_0^2)}{[(\omega_0^2-\omega^2)^2+
(\omega 
\gamma)^2]}=\frac{1}{4}
\frac
{(qE_0^{\prime})^2}{m}g(\omega)\end{equation}where \begin{equation}g(\omega)=\frac{(\omega^2+\omega_0^2)}{[(\omega_0^2-\omega^2)^2+
(\omega 
\gamma)^2]}.\end{equation} Equation (2) shows the relationship between the incident electric field and the field acting on 
an electron. With the variation of frequency in the ultraviolet we can imagine  that the amplitude of the incident field is kept constant. However, the amplitude
$E_0^{\prime}$ is dependent on $\chi$ which is frequency dependent.  Lorentz theory  is based on the
assumption that the response  $\chi$ of the medium to the external field is small [13]. In  equation (2), $\chi$
appears both in the numerator as well as in the denominator. With the depolarization factor $D$ positive, any variation of $\chi$  in the numerator will be offset to some extent by the variation in $\chi$ in the denominator. So we neglect the variation of the term
$E_0^{\prime}$ with frequency and assume it to be constant. To find the derivative of the average energy per cycle $\overline {E(\omega)}$, it is
sufficient to find the derivative of the function $g(\omega)$ given by  equation (21) with respect to $\omega$.
 Equating the derivative  to zero, we  find the condition  at 
which the 
average energy per cycle is maximum. It turns out that the frequency is 
given by 
 \begin{equation}
\omega=\omega_0[\sqrt{4-(\frac{\gamma}{\omega_0})^2}-1]^{\frac{1}{2}}.
\end{equation}If the incident electromagnetic wave can oscillate the bound
electrons steadily at  frequency $\omega$ given by the last equation, then the wave has to deliver 
maximum energy per cycle and its absorption will  be  maximum.

It is clear from  equation (22) that for real values of $\omega$ we should have the 
ratio $f=\frac{\gamma}{\omega_0}<\sqrt3$. By trial we take several positive 
values of $f$ upto its maximum
 of $\sqrt3$. For each value of $f$ we find the values of $\gamma$ and 
$\omega$ using  equation (22) with the known value of $\omega_0$. With these values we determine the refractive index $n$
and the extinction coefficient $\kappa$ in the resonance region using equations (A.3) 
and (A.4) respectively. We put these values in  equation (19) and try to see for
which value of $f$ this equation is satisfied. From $f$ which 
satisfies  equation (19) we  calculate 
$\gamma_t$ and  $\omega_t$ using  equation (22).

\section {Determination of the damping coefficient and the true resonance frequency}
\noindent
The results of an experiment performed with a prism made of  flint glass have been reported by
Chakrabarti [1]. In this experiment  $\frac{1}{n^2-1}$ has been plotted against the  inverse
wavelength squared at optical range following  equation (8). From this plot we have determined  the values of $\omega_n$, $\omega_p.$ 
The value of $\omega_0$ has been estimated  using equation (7). The errors in these frequencies are less than $1\%.$ Refractive indices as a function of wavelengths have been shown in table 1 and the parameters needed for this work have been shown in table 2.

It has been shown in the discussion following  equation (22) that the maximum value of
$f=\frac{\gamma}{\omega_0}$ is $\sqrt3$. So we take trial values of $f$ like
0.1, 0.2 upto 1.7. For all these values of $f$ we determine $\gamma$
and then $\omega$ using
equation (22). We then determine $n$ and $\kappa$ using equations (A.3) and (A.4) 
respectively and calculate $A$,$B$,$C$ 
according to equations (15),(16),(17) respectively. With these values we try to see whether they
satisfy equation (19) or not.
The algorithm for this is as follows: suppose we call the left side of
 equation (19), $Y1$ and the right side, $Y2$. Now we form a parameter,
 $Y=Y1-Y2$.
For all values of $f$ from 0.1 to 1.7 we calculate $Y$ and find between which 
two values of $f$, the sign of $Y$ changes. In our case there was only one sign 
change in $Y$ between  $f$ values 0.6 and 0.7. Now we check this region 
more closely, that
is  between 0.60 to 0.70 at an interval of 0.01. We find that for
$f=0.65$, $Y=-0.001$ and for $f=0.66$, $Y=0.053$. So there is a zero crossing
between $f$ values 0.65 and 0.66. Proceeding similarly we finally find that
for $f=0.6501$, $Y=-0.0005$ and for $f=0.6502$, $Y=0.000055$. So the
solution lies between 0.6501 and 0.6502. We take the 
solution as $f=0.65015$ with an error of $0.00005$ which is just $\pm 0.008\%$. From this value of  $f$ we get $\gamma_t$ since we know the value of $\omega_0$. We get the value of $\omega_t$ by putting the value of $f$ in  equation (22). These values 
 are shown in table 3. The values of the  damping coefficient  and the true resonance frequency turn out to be $$\gamma_t= 11.6\times 10^{15}\text{
rad/s}$$ with an error of only $0.76\%$ and $$\omega_t=16.8\times 10^{15}\text {rad/s}$$  with an error  similar to the error in $\omega_0$. $\omega_t$ lies between $\omega_n$ and $\omega_0$. This solution is unique.

We find that $\omega_t$ differs significantly from  $\omega_n$. We have presented the parameters at true resonance frequency in table 4.  The maximum value of the absorption coefficient $\alpha$ at the true resonance frequency has come out to be $8.18\times 10^7 \text m^{-1}$ corresponding to an attenuation length
$12.2$ nm.

Once we determine the damping coefficient, we can determine the absorption coeffficient $\alpha$ in a small frequency range about $\omega_t$ where we assume $\gamma$ is constant and equal to $\gamma_t$. In figure 1 we plot $\alpha$ as a function of $x=\frac{\omega}{\omega_0}$ between $x=\frac{\omega_n}{\omega_0}=0.81$ to 1. We assume that within the frequency range $\omega_n$ to $\omega_0$, $\gamma$ remains constant. We clearly find that $\alpha$ attains a maximum value at $x=\frac {\omega_t}{\omega_0}=0.94.$ Jackson [14]
 has given the frequency
range for absorption coefficient $\alpha$ in the ultraviolet as well as the plasma
frequency of water. In the ultraviolet, we are concerned with
electronic oscillations as the most important component responsible
for dispersion. So the properties of glass will not be too different
from water at these frequencies. This work [14] shows that the maximum value of $\alpha$ is around $10^8 m^{-1}$ which is approximately the same as the maximum value that we have obtained. The frequency at which the maximum occurs is also the same in order of magnitude  as  $\omega_t.$ The full width at half maximum of the absorption curve can be read off approximately. It is determined to be approximately $15\times 10^{15}\text {rad/s}$ which is of the same order of magnitude that we have obtained for the value of $\gamma_t$. 

The damping coefficient determined in this work  is rather large.This broad absorption in the ultraviolet is due to the outer electrons in the atoms and molecules of the solid which take part in the process of dispersion.  The outer electrons are affected by the collisions and the electric fields of the neighbouring atoms. Consequently  an extensive region of continuous absorption is obtained in solids and liquids [8,15]. So this large value of $\gamma_t$ is expected. This value of $\gamma_t$ is valid only in the resonance region.

\section {Some other  results  }
\noindent
We assume the damping coeffficient $\gamma_t$ is constant within
 the small frequency range from $\omega_n$ to $\omega_0$ in which $\omega_t$ lies.
 We now find the frequency $\omega_{\kappa}$ at which the extinction coefficient $\kappa$ is maximum, assuming $\omega_{\kappa}$  is close to $\omega_t$. 
 For $\kappa$ to be maximum,  $\frac {d\kappa}{d\omega}$ should be zero and from equation (14) this should occur when  $A=B.$ 
Here we scan a parameter $\frac{\omega}{\omega_0}$ from 0.1  to 1.0. For each value of $\omega$ and known $\gamma_t$ we calculate
$n$ and $\kappa$ using equations (A.3) and (A.4) respectively. We then check whether the condition $A=B$ is satisfied.
Once again we find a unique solution  with $\frac{\omega_{\kappa}}{\omega_0}=0.84843.$ This corresponds  to an  
angular  frequency  $\omega_{\kappa}=15.1\times 10^{15}\text{rad/s}$. From this we get 
$$\nu_{\kappa}=2.40\times 10^{15 }\text{Hz}$$  corresponding to a wavelength  $$\lambda_{\kappa}=0.125 \mu\text {m}.$$  In figure 2 we have plotted
 $\kappa$ determined by our theoretical analysis, as a function of $\frac{\omega}{\omega_0}.$ This figure clearly shows that $\kappa$ has attained a maximum value. The maximum value of $\kappa$ that we have obtained at $\omega=\omega_{\kappa}$ is 0.769. This is of the same order of magnitude as the maximum value  shown in 
 figure 2 of Kitamura et al.[3] which  gives the maximum of $\kappa$ in the ultraviolet at a wavelength very near to $0.12 \mu $m. This is very close to the wavelength $\lambda_{\kappa}$ that we have obtained.  In the last column of table 3 we show the value of $\omega_{\kappa}$. $\lambda_{\kappa}$ that we have determined  may be seen  from figure 1 of [3] to be the nearest to the optical wavelengths on long wavelength side as we claimed in last paragraph of section 2. The data in figure 1 of [3] may be showing other resonance/s at shorter wavelengths or larger frequencies.

In figure 3 we show a plot of reflectance $R$ as a function of frequency in the absorption region where $R$ is the normal reflectivity defined in  [3] as \begin{equation}R= \frac {(n-1)^2+\kappa^2}{(n+1)^2+\kappa^2}.\end{equation} We find that this function has a peak at $\frac {\omega}{\omega_0}=0.908$ with the maximum value of $R=0.118$. The peak position corresponds to $\omega=16.2 \times 10^{15}\text{rad/s}. $ The reflectance peak position corresponds to an energy 10.6 eV. Experimentally, silica glasses show the lowest frequency  reflectance peak in the ultraviolet region  at 10.2 eV [16] which is very close to what we have found theoretically. This is once again as we claimed in the last paragraph of section 2. Sigel  has shown in figure 3 [16] that the reflectance peak position is the same for two glassy materials though their reflectance values at this frequency are  different. Similarly the actual values of refractive indices and  extinction coefficients  at the same  frequency can vary significantly due to glass manufacturing process [3]. Data for these coefficients in the absorption region for flint glasses were not available in the literature. However, the order of magnitude of these coefficients is similar to that of  other silica glasses. Since we are getting the peak positions of the extinction coefficient and the reflectance data very close to the published data for silica glasses, we  conclude that the value of the damping coefficient obtained by us is correct.

We have shown in table 5 the  values of $n$, $\kappa$, $\alpha$, the absorption length 
$ \frac{1}{\alpha}$ and $g(\omega)$ at frequencies close to $\omega_t$. They have been calculated at three
 frequencies  $\omega_0$, $\omega_{\kappa}$  and $\omega_n$ for the same $\gamma_t$.
  Comparing tables 4 and 5 we find that $\alpha$  and $g(\omega)$ are  indeed maxima at  
  $\omega_t$.
The anomalous nature of variation of the refractive index is evident 
from the values of $n$ at frequencies around $\omega_t$. 

In section 4 we determined the condition for $\alpha$ to be maximum. In a similar way we can find the condition for the refractive index $n$ to be an extremum by equating its derivative to zero.  This derivative can be obtained by eliminating $\frac {d\kappa}{d\omega}$ from  equations (B.2) and (B.3). Interestingly, this condition  gives two solutions for the frequency. The refractive index is  maximum at one  and  minimum at  the other frequency. We find that the Lorentz-Maxwell model of dispersion reproduces all the features of anomalous dispersion in the absorption region as observed in actual experiments [3].

\section { Conclusions}
\noindent
The Lorentz-Maxwell model of dispersion of electromagnetic waves  in matter has been studied in this paper with an analysis of the phenomenon of  absorption in the ultraviolet  in dielectrics like flint glass. We have shown that if we know the refractive indices of glass fairly accurately in the optical frequencies, we can explore the anomalous dispersion region in the ultraviolet quantitatively. The key finding of this work is that the damping coefficient of the model   can be determined by a simple argument. We also determine the frequency at which the absorption coefficient is maximum. We call this the true resonance frequency.  In the optical region where glass is transparent, the damping coefficient can be assumed to be zero. In the absorptive part the damping coefficient has been taken to be a constant within a short range of frequencies. The value of the damping coefficient matches in order of magnitude with the experimental  width of the  absorption coefficient data for water available from literature. 

Once the damping coefficient is determined, we can find the frequencies at which the extinction coefficient and the reflectance  are maxima. These frequencies  match very well with the experimental data available in the literature for silica glasses. This indirectly shows that the value of $\gamma$ determined by us is correct. Our assumption of a single resonance should give us the information of the absorption band closest to the optical frequencies. We actually observe this by comparing  the peak positions of the extinction coefficient and reflectance data  obtained by us  with that found from  literature.
 Refractive indices estimated at different frequencies close to the true resonance frequency in the absorption region reveal the anomalous nature of dispersion. All the features of  dispersion by a  dielectric like glass in the ultraviolet absorption region  have been reproduced  from our theoretical analysis of the Lorentz-Maxwell model.

\section{Acknowledgment}
\noindent

The author would like to thank Prof. Jayanta Kumar Bhattacharjee for some helpful discussions.
 Thanks are due to Prof. Debashis Mukherjee  for making some helpful comments on the paper.
 
\newpage
\appendix

\renewcommand{\theequation}{A.{\arabic{equation}}}

\setcounter{equation}{0}

\begin{center}

\large{\bf Appendix A} \\
\end{center}
\noindent
Equations  (11) and (12) can be easily inverted and we get [2],
\begin{equation}n^2=\frac{1}{2}[(\epsilon_1^2+\epsilon_2^2)^{\frac{1}{2}}+
\epsilon_1]\end{equation} and \begin{equation}\kappa^2=\frac{1}{2}[(\epsilon_1^2+
\epsilon_2^2)^{\frac{1}{2}}-
\epsilon_1].\end{equation}
We  express $\epsilon_1$ and $\epsilon_2$
as functions of frequency using  equations (11) and (12) respectively. After a fairly
straightforward algebra we arrive at the  final
expressions for $n^2$ and $\kappa^2$ as functions of frequency $\omega$.

\begin{equation}
n^2 = \frac{[\omega_p^4\gamma^2\omega^2 + ({\omega^\prime}^4+\gamma^2\omega^2
+\omega_p^2{\omega^\prime}^2)^2]^{\frac{1}{2}}+({\omega^\prime}^4+\gamma^2
\omega^2+\omega_p^2{\omega^\prime}^2)}{2({\omega^\prime}^4+\gamma^2\omega^2)}
\end{equation} \begin{equation}
\kappa^2=\frac{[\omega_p^4\gamma^2\omega^2 + ({\omega^\prime}^4+\gamma^2\omega^2
+\omega_p^2{\omega^\prime}^2)^2]^{\frac{1}{2}}-({\omega^\prime}^4+\gamma^2
\omega^2+\omega_p^2{\omega^\prime}^2)}{2({\omega^\prime}^4+\gamma^2\omega^2)}
\end{equation}
where \begin{equation}
{\omega^\prime}^2=\omega_n^2-\omega^2.\end{equation}
These equations are exact and will be used for determining $n$ and $\kappa$
in the resonance region in the ultraviolet. 

It can be easily checked that in the limit  $\gamma$ tending to zero,  $\kappa$ becomes zero  at all frequencies and vice versa. Thus the medium is transparent  in the optical frequencies as it should. In this limit the refractive index $n$  satisfies a relation which has been used  in the first place  to get the parameters $\omega_n$ and $\omega_p$ [1]. 

\renewcommand{\theequation}{B.{\arabic{equation}}}

\setcounter{equation}{0}
\begin{center}

\large{\bf Appendix B} \\
\end{center}
\noindent
We first find the  derivative of $\frac{1}{(\omega_n^2-\omega^2)^2+\gamma^2\omega^2}$ with respect to $\omega$ and get \begin{equation}\frac{d}{d\omega}[\frac{1}{(\omega_n^2-\omega^2)^2+\gamma^2\omega^2}]=\frac{2\omega[2(\omega_n^2-\omega^2)-\gamma^2]}{[(\omega_n^2-\omega^2)^2+\gamma^2\omega^2]^2}\end{equation}
Differentiating  equation (11) with respect to $\omega$ we get
\begin{equation}\begin{split}n\frac{dn}{d\omega}-\kappa\frac
{d\kappa}{d\omega}&=\frac{-\omega\omega_p^2[(\omega_n^2-\omega^2)^2+\gamma^2\omega^2]+\omega\omega_p^2(\omega_n^2-\omega^2)[2(\omega_n^2-\omega^2)-\gamma^2]}{[(\omega_n^2-\omega^2)^2+\gamma^2\omega^2]^2}\\
&=\frac{\omega\omega_p^2[(\omega_n^2-\omega^2)^2-
\omega_n^2\gamma^2]}{[(\omega_n^2-\omega^2)^2+\gamma^2 \omega^2]^2}\end{split}\end{equation} Similarly differentiating  equation (12) with respect to $\omega$ we get \begin{equation}\begin{split}2\kappa\frac{dn}{d\omega}+2n\frac
{d\kappa}{d\omega}&=\frac{\omega_p^2\gamma[(\omega_n^2-\omega^2)^2+\gamma^2\omega^2]+2\omega^2\omega_p^2
\gamma[2(\omega_n^2-\omega^2)-\gamma^2]}{[(\omega_n^2-\omega^2)^2+\gamma^2\omega^2]^2}\\&=
\frac{\omega_p^2\gamma (\omega_n^2-\omega^2)[(\omega_n^2-
\omega^2)+4\omega^2]-\omega^2\omega_p^2\gamma^3}{[(\omega_n^2-\omega^2)^2+\gamma^2\omega^2]^2}
\end{split}.\end{equation}

\newpage
\begin{center}
{\large\bf REFERENCES}
\end{center}
\begin{enumerate}
\item  Chakrabarti S 2006  Phys. Educ. { \bf 23 }167- 175 (New Delhi, India: South Asian Publishers PVT LTD)
\item  Christy R W 1972  Am.J.Phys. {\bf 40} 1403-1419
\item  Kitamura R, Pilon L and Jonasz M 2007   Applied Optics {\bf 46(33)} 8118-8133
\item Almog I F, Bradley M S  and Bulovic V 2011  The Lorentz Oscillator and its Applications ( MIT OpenCourseWare, MIT6-007S11/lorentz)
\item Born M, Wolf E 1980  \textit{Principles of Optics}  ( New York:Pergamon Press, Sixth.Edn.) p  85 
\item  Kittel C 1976 \textit{Introduction To Solid State Physics}  (New Delhi:WileyEastern Limited, 5th Edn.) p 405 
\item Hecht E  2002 \textit{Optics}  ( Delhi,India:Pearson Education, 4th Edn.)  p 71, 85,70,128
\item Feynman R P, Leighton R B, Sands M 2003 \textit{The Feynman Lectures on Physics, 2nd Volume} (New Delhi,India:Narosa Publishing House) p 1211,1210
\item Oughstun K E, Cartwright N A  2003 Optics Express { \bf 11(13)} 1541-1546
\item Tanner D B  2013  Optical effects in solids (www.phys.ufl.edu/tanner/notes.pdf )
\item Kleppner D, Kolenkow R J 1973  \textit{An Introduction To Mechanics} (New Delhi :Tata Mcgraw Hill )  p 426 
\item Heitler W 1954  \textit{The quantum theory of radiation} (Oxford University Press, 3rd. Edn.)  p 163 
\item Seitz F 1940  \textit{The Modern Theory of Solids} (McGraw-Hill ,International Series In Pure And Applied Physics)  p 629 
\item Jackson J D 1999  \textit{Classical Electrodynamics  3rd edn.}  (John Wiley $\&$ Sons, Inc) p 314 
\item Jenkins F A , White H E 1957  \textit{Fundamentals of Optics}  (Mcgraw-Hill book company,Inc, 3rd Edn.)  p 486 
\item Sigel G H  Jr  1973/74 Journal of Non-Crystalline Solids {\bf 13 }  372-398  
\end{enumerate}

\newpage
\begin {table}[h!]
\centering
\caption {\bf Refractive indices as a function of wavelengths for the  flint glass prism [1]}
\begin{tabular}{cc}\\
\hline
wavelength &refractive index \\
$\lambda$(nm)&$n$\\
\hline
706.544&1.6087\\
667.815&1.6108\\
587.574&1.6167\\
504.774&1.6259\\
501.567&1.6264\\
492.193&1.6277\\
471.314&1.6311\\
447.148&1.6358\\
438.793&1.6377\\
\hline
\end{tabular}
\end {table}

\begin {table}
\caption {\bf Parameters obtained from fitting of data of refractive indices to Lorentz model[1]}
\begin {center}
\begin{tabular}{cccc}
\hline
$\omega_n$&$N$&$\omega_p$&$\omega_0$\\
$\text {rad/s}$&$m^{-3}$&$\text {rad/s}$&$\text {rad/s}$\\
\hline
$14.5\times 10^{15}$&$1.02\times 10^{29}$&$18.0\times 10^{15}$&$17.8\times 10^{15}$\\
\hline
\end{tabular}
\end{center}
\end {table}

\begin {table}
\caption {\bf  Table for  $\gamma_t$ , $\omega_t$ and $\omega_{\kappa}$}
\begin {center}
\begin{tabular}{cccc}
\hline
$f=\frac{\gamma_t}{\omega_0}$&$\gamma_t$&$\omega_t$&$\omega_{\kappa}$\\
&$\text{rad/s}$&$\text{rad/s}$&$\text{rad/s}$\\
\hline
0.65015&$11.6\times10^{15}$&$16.8\times10^{15}$&$15.1\times10^{15}$\\
\hline
\end{tabular}
\end{center}
\end {table}

\begin {table}[h]
\caption {\bf Parameters at the true resonance frequency }
\begin {center}
\begin{tabular}{cccccc}
\hline
frequency&n&$\kappa$&$\alpha$&$\frac{1}{\alpha}$&$g(\omega)=\frac{4m\overline{E(\omega)}}
{(qE_0^{\prime})^2}$\\
$\omega$&&&$m^{-1}$&$\mu$m&$\text{s}^2/\text{rad}^2$\\
\hline
$\omega_t$&0.995&0.729&$8.18\times10^7$&0.0122&$0.153\times10^{-31}$\\
\hline
\end{tabular}
\end{center}
\end {table}
\begin {table}[h]
\caption {\bf Values of some parameters in the ultraviolet region}
\begin {center}
\begin{tabular}{cccccc}
\hline
frequency&n&$\kappa$&$\alpha$&$\frac{1}{\alpha}$&$g(\omega)=\frac{4m\overline{E(\omega)}}
{(qE_0^{\prime})^2}$\\
$\omega$&&&$m^{-1}$&$\mu$m&$\text{s}^2/\text{rad}^2$\\
\hline
$\omega_0$&0.906&0.678&$8.06\times10^7$&0.0124&$0.149\times10^{-31}$\\
$\omega_{\kappa}$&1.18&0.769&$7.76\times 10^7$&0.0129&$0.141\times 10^{-31}$\\
$\omega_n$&1.26&0.763&$7.39\times10^7$&0.0135&$0.133\times10^{-31}$\\
\hline
\end{tabular}
\end{center}
\end {table}

\begin{figure}[h!]
\centering
\includegraphics[width=14 cm]{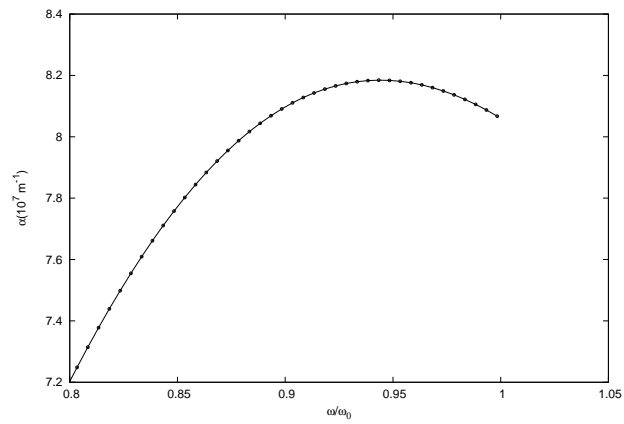}
\caption{Distribution of the absorption  coefficient $\alpha$ in the absorption region}
\end{figure}

\begin{figure}[h!]
\centering
\includegraphics[width=14 cm]{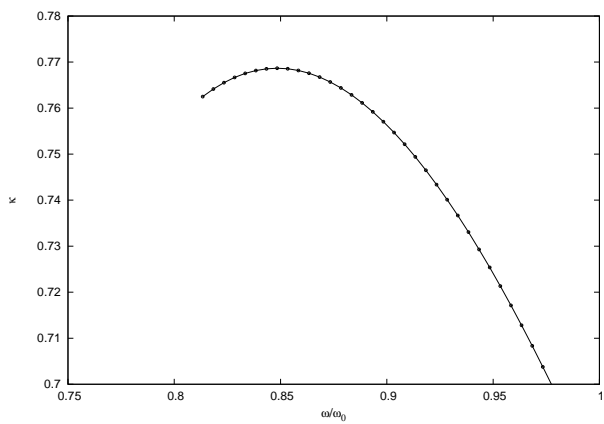}
\caption{Distribution of the extinction coefficient $\kappa$ in the absorption region}
\end{figure}

\begin{figure}[h!]
\centering
\includegraphics[width=16 cm]{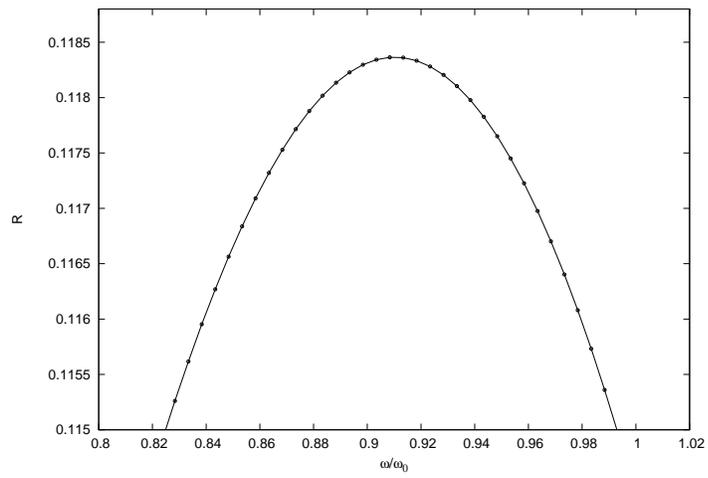}
\caption{Distribution of Reflectance $R$ in the absorption region}
\end{figure}
\end{document}